\documentclass[12pt,twoside,english]{article}
\usepackage[T1]{fontenc}
\usepackage[latin1]{inputenc}
\usepackage{color}
\usepackage{amsmath}
\usepackage{amssymb}



\usepackage{babel}
\begin{document}
\title{Macroscopic quantum-mechanical scattering}
\author{{\normalsize{}M. Apostol }\\
{\normalsize{}Department of Theoretical Physics, Institute of Atomic
Physics, }\\
{\normalsize{}Magurele-Bucharest MG-6, POBox MG-35, Romania }\\
{\normalsize{}email: apoma@theory.nipne.ro}}
\date{{}}

\maketitle
\relax
\begin{abstract}
In certain conditions a macroscopic quantum-mechanical scattering
may occur, which may lead to a coherent cross-section on a macroscopic
scale in a monocrystal. The conditions are satisfied by neutrinos,
but not satisfied by other projectiles, with a higher cross-section.
This may explain Weber-type experiments of neutrino detection by a
perfect, stiff sapphire monocrystal. The occurrence of coherence domains
for quantum-mechanical scattering and classical diffraction is analyzed,
and the force exerted upon a macroscopic target is estimated. It is
concluded that neutrinos exhibit a distinctive feature in this respect,
due precisely to their very small cross-section. 
\end{abstract}
\relax

Key words: neutrino scattering; coherent scattering; macroscopic coherence;
neutrino detection 

\noindent \textbf{Introduction.} In two papers published in 1985
and 1988 Weber claimed that neutrinos (antineutrinos), from various
sources like tritium, nuclear reactors and Sun, could be detected
by their coherent scattering by a perfect, stiff, sapphire monocrystal
with a high Debye temperature (mounted on a torsion balance and equilibrated
by a lead dummy).\cite{key-1,key-2} The coherent cross-section would
be $\sigma=N^{2}\sigma_{0}$, where $N$ is the number of unit cells
in the target and $\sigma_{0}$ is the cross-section of a single unit
cell (particle, \emph{e.g.}, atomic nucleus). Such a highly enhanced
cross-section $\sim N^{2}$ would give rise to a measurable force
upon a torsion balance. Weber's claims have been criticized both on
theoretical and experimental grounds, the main objection being that
the form-factor would reduce appreciably the cross-section, and, on
the other hand, such a coherence effect is not observed in $X-$,
gamma rays or neutron scattering\cite{key-3}-\cite{key-12} (see
also Refs. \cite{key-13,key-14}). A discussion of the theoretical
objections and negative experiments was given by Nicolescu, who presented
a positive experiment;\cite{key-15} indeed, an experiment by Cruceru
et al exists, which confirms Weber's prediction for solar neutrinos.\cite{key-15}-\cite{key-17}
The problem is still controversial. We show in this paper that a coherent
scattering of neutrinos may appear in the conditions formulated by
Weber and company, as a consequence of a quantum-mechanical treatment
of the crystal as a whole (a macroscopic quantum-mechanical scattering).
This is a distinctive condition of the neutrino scattering, which
is not fulfilled by other projectiles, with a higher cross-section.
The main reason for such a behaviour is precisely the extremely small
cross-section $\sigma_{0}$ (of the order $10^{-44}cm^{2}$) of the
neutrinos. 

For $\sigma_{0}=10^{-44}cm^{2}$ and $N=10^{22}$ (less than $0.1\,mol$)
the coherent cross-section is $\sigma=1cm^{2}$. For a neutrino flux
density $\Phi=10^{12}/cm^{2}\cdot s$ the time between two collisions
is $\tau=1/\Phi\sigma=10^{-12}s$. On the other hand, an atom in thermal
equilibrium at room temperature has a velocity of the order $v=10^{4}cm/s$.
In an elementary act of collision the atom is perturbed from its equilibrium
state and receives a momentum of the order $p=E/c$, on average, where
$E$ is the energy of the neutrino projectile (and $c=3\times10^{10}cm/s$
is the speed of light). For $E=1MeV$ the momentum transfer is of
the order $p=5\times10^{-17}g\cdot cm/s$. Consequently, the energy
perturbation of the atom is of the order $\Delta E=p^{2}/M=2.5\times10^{-11}erg$
for $M=10^{5}m_{e}$, where $m_{e}\simeq10^{-27}g$ is the electron
mass ($vp=5\times10^{-13}erg\ll p^{2}/2M$). The time needed for this
atom to recover its equilibrium is of the order $\Delta t_{eq}=\hbar/\Delta E\simeq4\times10^{-17}s$
(where $\hbar\simeq10^{-27}erg\cdot s$ is Planck's constant). We
can see that $\tau\gg\Delta t_{eq}$. It follows that the atoms recover
quickly their equilibrium state between two sucessive collisions,
and the incident neutrino beam sees the crystal as a whole. Therefore,
we need to adopt a quantum-mechanical treatment for the entire crystal.
It is worth noting that if the cross-section increases to, say, $\sigma_{0}=10^{-24}cm^{2}$,
as for $X$-, gamma rays or neutrons, the \textquotedbl collision\textquotedbl{}
time decreases to $\tau=10^{-32}s$, which is much shorter than the
equilibrium time $\Delta t_{eq}$, all the other conditions remaining
the same. In that case the incident projectile beam sees the crystal
as consisting of distinct, independent atoms, such that a coherent
scattering ($\sigma=N^{2}\sigma_{0}$) for the entire crystal is not
possible. The particularity of a coherent scattering suffered by neutrinos
in the whole crystal resides precisely in their extremely small cross-section
$\sigma_{0}$. On the other hand, a single-particle cross-section
$\sigma_{0}=10^{-24}cm^{2}$ increases considerably the total cross-section,
such that we need to reconsider the scattering in this case.

\textbf{Macroscopic quantum-mechanical scattering.} Let us assume
a macroscopic target consisting of $N\gg1$ identical \textquotedbl atoms\textquotedbl{}
(atomic nuclei, molecules, unit cells in a crystal). The interaction
with an incident beam of particles can be written as 
\begin{equation}
H=a^{3}h(\xi)\sum_{i=1}^{N}\delta(\boldsymbol{r}-\boldsymbol{r}_{i})\,\,\,,\label{1}
\end{equation}
 where $a$ is the range of the single-particle interaction $h(\xi)$,
$\xi$ denotes the internal coordinates of the atoms and $\boldsymbol{r}_{i}$
are the atomic positions. The time between two successive collisions
is $\tau=1/\Phi\sigma$, where $\Phi$ is the incident flux density
and $\sigma$ is the total cross-section. In an elementary act of
collision an atom receives a momentum of the order of the momentum
$p$ of the incident particle. The atom has a thermal velocity $v\simeq\sqrt{T/M}$,
where $T$ is the temperature and $M$ is the mass of the atom. The
atomic motion is perturbed by an energy of the order $\Delta E=p^{2}/M+vp$,
so it needs a time $\Delta t_{eq}\simeq\hbar/\Delta E$ to recover
its equilibrium. Let us assume 
\begin{equation}
\tau>\Delta t_{eq}\,\,;\label{2}
\end{equation}
then, the incident particles see the macroscopic target as a whole,
and we need to work with the wavefunction of the whole, macroscopic,
target. 

A perfect monocrystal suffers two kinds of motion. One kind is the
motion of the crystal as a whole, where all the laticial positions
of the atoms $\boldsymbol{r}_{i}^{0}$ move by the same distance.
The wavefunction corresponding to this motion is 
\begin{equation}
\Phi_{\boldsymbol{K}}(\xi;\boldsymbol{r})=\sum_{i=1}^{N}e^{i\boldsymbol{K}\boldsymbol{r}_{i}^{0}}\varphi(\xi;\boldsymbol{r}-\boldsymbol{r}_{i}^{0})\,\,\,,\label{3}
\end{equation}
 where $\boldsymbol{K}$ is the quasi-wavevector of the crystal (quasi-momentum
$\hbar\boldsymbol{K}$) and $\varphi(\xi;\boldsymbol{r}-\boldsymbol{r}_{i}^{0})$
are (orthonormal) wavefuctions localized on the positions $\boldsymbol{r}_{i}^{0}$.
We can see that the wavefunction $\Phi_{\boldsymbol{K}}$ has the
translational symmetry of the crystal. We call $\Phi_{\boldsymbol{K}}$
a coherent wavefunction. The other type of motion of the crystal is
the thermal motion with atomic displacements $\boldsymbol{v}_{i}$
($\boldsymbol{r}_{i}=\boldsymbol{r}_{i}^{0}+\boldsymbol{v}_{i})$;
the corresponding wavefunction is 
\begin{equation}
\psi(\xi;\boldsymbol{r})=\sum_{i=1}^{N}e^{i\chi_{i}}\varphi(\xi;\boldsymbol{r}-\boldsymbol{r}_{i}^{0}-\boldsymbol{v}_{i})\,\,\,,\label{4}
\end{equation}
 where $e^{i\chi_{i}}$ are random phase factors; we call $\psi$
an incoherent wavefunction. Both these wavefunctions are normalized
to $N$; they are orthogonal to each other. The wavefunction of the
crystal is 
\begin{equation}
\varPsi_{\boldsymbol{K}}=\sqrt{1-g^{2}}\Phi_{\boldsymbol{K}}+g\psi\,\,\,,\label{5}
\end{equation}
 where $g$ is a weight coefficient. This coefficient is proportional
to the square root of the relative number of vibration states of the
crystal, properly normalized. This relative number of phonon states
is proportional to 
\begin{equation}
I=\frac{1}{\omega_{D}^{3}}\int_{0}^{\omega_{D}}d\omega\frac{\omega^{2}}{e^{\hbar\omega/T}-1}\,\,\,,\label{6}
\end{equation}
where $\omega_{D}$ is the Debye frequency and $\omega$ is the phonon
frequency; the Debye temperature is $\Theta=\hbar\omega_{D}$. For
$T/\Theta\ll1$ the integral is $I\simeq2.4(T/\Theta)^{3}$, while
for $T/\Theta\gg1$ the integral is $I\simeq\frac{1}{2}T/\Theta$.
A normalized expression for the relative number of states is $4.8(T/\Theta)^{2}$
for $T/\Theta\ll1$ and $1$ for $T/\Theta\gg1$. An interpolation
formula is $I\simeq4.8(T/\omega_{D})^{2}/\left[1+4.8(T/\omega_{D})^{2}\right]$,
such that we can take for the weight coefficient 
\begin{equation}
g\simeq\frac{2.2(T/\Theta)}{\left[1+4.8(T/\Theta)^{2}\right]^{1/2}}\,\,.\label{7}
\end{equation}
We can see that for high Debye temperatures the main contribution
to $\varPsi_{\boldsymbol{K}}$ comes from the coherent wavefunction
($g\ll1$), while for low Debye temperatures the main contribution
comes from the incoherent field ($g\rightarrow1$). 

Let us assume a wavefunction $\frac{1}{\sqrt{V}}e^{i\boldsymbol{kr}}$
for an incident particle, and an initial ($i$) wavefunction $\frac{1}{\sqrt{V}}e^{i\boldsymbol{kr}}\varPsi_{\boldsymbol{K}}$,
where $V$ denotes the volume. The normalization to unity of this
wavefunction requires a factor $\sqrt{a^{3}}$ in the wavefuctions
$\varphi$, such that the scalar product is $\left\langle \varphi(\xi;\boldsymbol{r}_{j}-\boldsymbol{r}_{i}),\varphi(\xi;\boldsymbol{r}_{j}-\boldsymbol{r}_{i})\right\rangle ^{2}=\delta_{ij}$.
The matrix elements of the interaction between the wavefunctions $\Phi_{\boldsymbol{K}}$
and $\psi$ are zero. The matrix elements of the interaction between
two wavefunctions $\Phi_{\boldsymbol{K}}$ and $\Phi_{\boldsymbol{K}'}$
(coherent scattering) lead to the momentum conservation 
\begin{equation}
H_{fi}\sim\sum_{i=1}^{N}e^{i(\boldsymbol{K}-\boldsymbol{K}')\boldsymbol{r}_{i}^{0}}e^{i(\boldsymbol{k}-\boldsymbol{k}')\boldsymbol{r}_{i}^{0}}=N\delta_{\boldsymbol{K}'+\boldsymbol{k}',\boldsymbol{K}+\boldsymbol{k}}\,\,\,,\label{8}
\end{equation}
 where $\boldsymbol{k}',\,\boldsymbol{K}'$ are the wavevectors of
the final state ($f$). We can see that the difference in momentum
of the incident particle is taken by the crystal, which moves as a
whole. According to equation (\ref{8}) the coherent cross-section
is 
\begin{equation}
\sigma_{coh}=N^{2}\sigma_{0}\,\,\,,\label{9}
\end{equation}
 where $\sigma_{0}$ is the single-particle cross-section. A similar
calculation for the incoherent matrix elements (wavefunctions $\psi$)
leads to 
\begin{equation}
H_{fi}\sim\sum_{i=1}^{N}e^{i(\boldsymbol{k}-\boldsymbol{k}')\boldsymbol{r}_{i}^{0}}\left(1+i(\boldsymbol{k}-\boldsymbol{k}')\boldsymbol{v}_{i}+...\right)\,\,\,,\label{10}
\end{equation}
 where 
\begin{equation}
\boldsymbol{v}_{i}=\frac{1}{\sqrt{N}}\sum_{\boldsymbol{q}}e^{i\boldsymbol{q}\boldsymbol{r}_{\boldsymbol{i}}^{\boldsymbol{0}}}\boldsymbol{v}_{\boldsymbol{q}}\label{11}
\end{equation}
 is the phonon field. The first term in equation (\ref{10}) corresponds
to a displacement of the crystal as a whole, so it is already included
in the coherent scattering. We are left with 
\begin{equation}
H_{fi}\sim\sqrt{N}\left(\boldsymbol{q}\boldsymbol{v}_{\boldsymbol{q}}\right)\delta_{\boldsymbol{k}',\boldsymbol{k}+\boldsymbol{q}}\,\,.\label{12}
\end{equation}
 We can see that the difference in momentum of the incident particle
is taken by phonons; the incoherent scattering excites phonons. In
addition, the incoherent cross-section is proportional to $N$. For
the cross-section we need to average $(\boldsymbol{q}\boldsymbol{v}_{\boldsymbol{q}})^{2}$
over the thermal states. The maximum value of this average is of the
order $T/Mc_{s}^{2}$ (at room temperature), where $c_{s}$ is the
mean phonon velocity (of the order $10^{6}cm/s$). This thermal factor
is reminiscent of the Debye-Waller factor (and the diffuse scattering).
The total incoherent cross-section can be written as 
\begin{equation}
\sigma_{incoh}=N\sigma_{0}^{ph}\,\,\,,\label{13}
\end{equation}
 where the single-particle cross-section $\sigma_{0}^{ph}$, arising
from phonons, is smaller than $\sigma_{0}$ by the thermal factor. 

The total cross-section of the crystal is 
\begin{equation}
\sigma=\sqrt{1-g^{2}}\sigma_{coh}+g^{2}\sigma_{incoh}\,\,.\label{14}
\end{equation}
The incoherent cross-section is extremely small in comparison with
the coherent cross-section. For atoms placed randomly (like in amorphous
solids, liquids, etc) the coherent wavefunction $\Phi_{\boldsymbol{K}}$
is absent and the weight coefficient is $g=1$. 

\textbf{Neutrino scattering.} We adopt $N\simeq10^{22}$ for the number
of unit cells in the sapphire monocrystal ($\simeq24g$, density $4g/cm^{3}$)
used in Weber's experiments,\cite{key-1,key-2} and other similar
experiments\cite{key-15}-\cite{key-17} (the volume of the unit cell
of sapphire is large, $\simeq10^{3}\textrm{Å}^{3}$). Making use of
$\sigma_{0}=10^{-44}cm^{2}$ we get a coherent cross-section $\sigma_{coh}\simeq1cm^{2}$.
For a Debye temperature $\Theta=10^{3}K$ the weight factor at room
temperature is $\sqrt{1-g^{2}}\simeq0.7$. The total coherent cross-section
is $\sigma\simeq0.7cm^{2}$. We note that this cross-section is smaller
than the cross-sectional area of the crystal. For a flux density $\Phi=10^{12}/cm^{2}\cdot s$
the collision time is $\tau=1/\Phi\sigma_{coh}\simeq10^{-12}s$. At
room temperature the thermal velocity of an atom is $v\simeq10^{4}cm/s$.
For a neutrino energy $E=1MeV$ the momentum is $p=E/c\simeq5\cdot10^{-17}g\cdot cm/s$.
The perturbation energy is $\Delta E=p^{2}/M\simeq2.5\times10^{-11}erg$
(for an atomic mass $M=10^{5}m_{e}$, where $m_{e}$ is the electron
mass); the contribution $\Delta E=vp\simeq5\times10^{-13}erg$ is
much smaller. The equilibrium time is of the order $\Delta t_{eq}=\hbar/\Delta E\simeq4\times10^{-17}s$.
Since $\tau\gg\Delta t_{eq}$ the macroscopic quantum-scattering described
above applies. 

The force acting upon the target in the forward direction is $F=\Phi\sigma p\simeq3.5\times10^{-5}\,dyn$.
This is a measurable force. We note that it is sensitive to the values
of the input parameters. For instance, a Debye temperature $\Theta=100K$
leads to a weight coefficient $\sqrt{1-g^{2}}\simeq0.15$ and a weaker
force by a factor $\simeq5$. Also, for an amorphous solid, although
the conditions of a quantum-mechanical scattering may be fulfilled,
the force is extremely weak, as a consequence of the very small incoherent
cross-section. 

For solar neutrinos ($E\simeq300keV$) the single-particle cross-section
may preserve its value, but the flux density is smaller ($\Phi\simeq10^{11}/cm^{2}\cdot s$).
The conditions of a coherent scattering are preserved, but the force
may be diminished by a factor $\simeq10^{-1}$ ($\simeq10^{-6}dyn$).
Also, a decrease may appear for tritium neutrinos ($E=10keV,$ $\Phi\simeq10^{14}/cm^{2}\cdot s$),
though a higher $\sigma_{0}$ or a slightly greater number of unit
cells $N$ may compensate the decrease (while preserving the conditions
of coherent scattering). We note that a large number of atoms in the
unit cell, as for a sapphire crystal, may increase the single-particle
cross-section $\sigma_{0}$. We conclude that Weber-type experiments
could exhibit a measurable force acting upon a sapphire crystal. 

\textbf{Other projectiles. Coherence domains.} We adopt the value
$\sigma_{0}=10^{-24}cm^{2}$ for other types of projectiles (like
$X$-, gamma rays or neutrons). A coherent cross-section would be
much larger than the cross-sectional area of the crystal. The crystal
responds to this unphysical situation by developing coherence domains.
Let us assume that $n_{d}$ uncorrelated domains exist in the crystal,
each with $N_{d}$ unit cells (as a mean size), such that $n_{d}=N/N_{d}$.
By a formal analogy with the high-purity crystals we use the fraction
$f=1/N_{d}$. This fraction varies between $f=1/N$, when we have
only one domain, \emph{i.e.} the whole target, and $f=1$, when the
whole target is fragmented in \textquotedbl atomic\textquotedbl{}
domains. 

The scattering amplitude can be written as 
\begin{equation}
S=\sum_{a=1}^{n_{d}}e^{i\chi_{a}}\left(H_{fi}\right)_{a}\,\,\,,\label{15}
\end{equation}
 where $e^{i\chi_{a}}$ are random phase factors. By averaging the
squared scattering amplitude over the phase factors, we get $\mid S\mid^{2}=\sum_{a}\left|\left(H_{fi}\right)_{a}\right|^{2}$,
such that the cross-section becomes 
\begin{equation}
\overline{\sigma}=n_{d}\sigma_{d}\,\,\,,\label{16}
\end{equation}
 where $\sigma_{d}$ is the cross-section of a domain. 

According to this equation, the coherent cross-section $\sigma_{coh}=\sigma_{0}N^{2}$
is reduced by the coherence domains to 
\begin{equation}
\overline{\sigma}=\sigma_{0}n_{d}N_{d}^{2}=\frac{\sigma_{0}N}{f}\,\,;\label{17}
\end{equation}
we can see that this formula gives the total coherent cross-section
($\sigma_{0}N^{2}$) for $f=1/N$ and an incoherent cross-section
($\sigma_{0}N$) for $f=1$. In this latter case $\sigma_{0}$ should
be replaced by $\sigma_{0}^{ph}$ (a similar procedure leaves the
incoherent cross-section arising from phonons unchanged, $\overline{\sigma}_{incoh}=n_{d}N_{d}\sigma_{0}^{ph}=N\sigma_{0}^{ph}=\sigma_{incoh}$). 

In order to have a quantum-mechanical scattering the conditions $\overline{\tau}>\Delta t_{eq}$
and $\overline{\sigma}<A$ should be satisfied, where $A$ is the
cross-sectional area of the target; these conditions lead to 
\begin{equation}
f>\frac{\hbar\Phi}{\Delta E}\sigma_{0}N\,\,,\,\,f>\frac{\sigma_{0}N}{A}\label{18}
\end{equation}
and a number of unit cells $N_{d}=1/f<A/\sigma_{0}N$ in each domain.
For $\sigma_{0}=10^{-24}cm^{2}$ this number is too small for any
macroscopic target ($N_{d}<10^{2}A$, $N=10^{22}$); the domains are
not well defined, such that $f$ approaches unity and the scattering
tends to an incoherent scattering. We conclude that the quantum-mechanical
scattering cannot appear for large single-particle cross-sections,
like $\sigma_{0}=10^{-24}cm^{2}$. We note that for neutrinos ($\sigma_{0}=10^{-44}cm^{2}$)
$f>10^{-22}/A$, $N_{d}<10^{22}A$ and we may have one domain in the
whole target. The coherent scattering occurs for neutrinos in a crystal
precisely due to the small neutrino cross-section$\sigma_{0}.$ The
above considerations apply also to a polycrystalline target, where
$f$ is limited, in addition, by the size of the crystallites and,
consequently, the cross-section is much diminished.

\textbf{Classical scattering.} If inequation (\ref{2}) is not satisfied
(\emph{i.e.}, if $\tau<\Delta t_{eq}$), the incident particles see
the target \textquotedbl atoms\textquotedbl{} (atomic nuclei, molecules,
unit cells) as independent scatterers. We call this scattering a classical
scattering. We can see that this condition implies low energies. The
initial wavefunction is $\frac{1}{\sqrt{V}}e^{i\boldsymbol{k}\boldsymbol{R}}\frac{1}{\sqrt{V}}e^{i\boldsymbol{K}\boldsymbol{R}_{c}}$
(up to wavefunctions corresponding to the internal degrees of freedom),
where $\boldsymbol{k}$ is the wavevector of the incident particle,
$\boldsymbol{K}$ is the wavevector of the center of mass, $\boldsymbol{R}_{c}$
is the position of the center of mass and $\boldsymbol{R}=\boldsymbol{r}+\boldsymbol{R}_{c}$.
The matrix elements of the interaction given by equation (\ref{1}),
\begin{equation}
H_{fi}\sim\sum_{i=1}^{N}e^{i(\boldsymbol{k}-\boldsymbol{k}')\boldsymbol{r}_{i}}\delta_{\boldsymbol{K}'+\boldsymbol{k}',\boldsymbol{K}+\boldsymbol{k}}\,\,\,,\label{19}
\end{equation}
includes the form-factor 
\begin{equation}
F(\boldsymbol{k}-\boldsymbol{k}')=\sum_{i=1}^{N}e^{i(\boldsymbol{k}-\boldsymbol{k}')\boldsymbol{r}_{i}}\,\,\,,\label{20}
\end{equation}
where $\boldsymbol{k}'$ is the wavevector of the scattered particle
and $\boldsymbol{K}'$ is the final wavevector of the center of mass.
We can see that the total momentum, including the momentum of the
center of mass, is conserved. For a crystal $F(\boldsymbol{k}-\boldsymbol{k}')=N\delta_{\boldsymbol{k}',\boldsymbol{k}+\boldsymbol{g}}$,
where $\boldsymbol{g}$ is a reciprocal vector of the lattice. For
an amorphous target $\boldsymbol{g}=0$. It follows that we have diffraction
peaks. For the cross-section of a peak we get $d\sigma_{\boldsymbol{g}}\sim N^{2}do_{\boldsymbol{g}}$,
where the solid angle $o_{\boldsymbol{g}}$ extends to the range $\Delta o_{\boldsymbol{g}}\simeq N^{-2/3}(2\pi/dk')^{2}$,
where $d$ is the mean distance between unit cells (scatterers). It
follows $\sigma_{\boldsymbol{g}}\sim N^{4/3}/(dk')^{2}$. On the other
hand the number of peaks is $\simeq(dk')^{2}\gg1$, such that the
total cross-section is 
\begin{equation}
\sigma=N^{4/3}\sigma_{0}\,\,\,,\label{21}
\end{equation}
where $\sigma_{0}$ is the single-particle cross-section. For one
peak $\sigma$ should be divided by the number of peaks. As it is
well known, this cross-section is affected by the Debye-Waller factor
and diffuse scattering. According to equation (\ref{15}) for $n_{d}=N/N_{d}=fN$
domains the cross-section is 
\begin{equation}
\overline{\sigma}=n_{d}N_{d}^{4/3}\sigma_{0}=\frac{\sigma_{0}N}{f^{1/3}}\,\,.\label{22}
\end{equation}
 For $f=1/N$ we recover the total cross-section $N^{4/3}\sigma_{0}$
of one domain, while for $f=1$ the cross-section reduces to $N\sigma_{0}$
of an incoherent scattering. 

The conditions $\overline{\tau}=1/\Phi\overline{\sigma}<\Delta t_{eq}$
and $\overline{\sigma}<A$ lead to 
\begin{equation}
\frac{\sigma_{0}N}{A}<f^{1/3}<\frac{\hbar\Phi}{\Delta E}\sigma_{0}N\,\,\,,\label{23}
\end{equation}
 which implies $N_{d}=1/f<(A/\sigma_{0}N)^{3}$ ($\overline{\sigma}<A$).
For $\sigma_{0}=10^{-24}cm^{2}$ and $N=10^{22}$ the number of unit
cells $N_{d}<10^{6}A^{3}$ in a domain may indicate well-defined domains
for macroscopic targets. The force is bounded from above according
to the inequality $F<\Phi Ap$. The inequations (\ref{23}) imply
$\Delta E<\hbar\Phi A$, \emph{i.e.} $\Delta E<10^{-15}A\,erg$ ($\Phi=10^{12}/cm^{2}\cdot s$).
Therefore, we may set $\Delta E=vp$ and $p<\frac{\hbar\Phi}{v}A$.
Consequently, this upper bound is given by $F<\frac{\hbar}{v}(\Phi A)^{2}\simeq10^{-7}A^{2}dyn$
($v=10^{4}cm/s$). For any reasonably large area $A$ and flux density
$\Phi$ it is difficult to satisfy these conditions ($p<\hbar\Phi A/v$)
and to measure such a force in current experimental situations. For
higher energies the energy transfer is higher and we need to apply
inequations (\ref{18}); the large cross-section in this case leads
to an incoherent scattering. 

\textbf{Concluding remarks.} A quantum-mechanical scattering is identified
in certain conditions in macroscopic targets, which may lead to a
coherent cross-section in high-purity, stiff monocrystals. This coherent
scattering may explain the Weber-type experiments of neutrino detection
by using sapphire monocrystals. The coherent-scattering conditions
are not fulfilled by other types of projectiles, with a higher single-particle
cross-sections (like $X$-, gamma or neutrons). In these cases a classical
diffraction may occur in crystals, which generates a weak force, at
the limit of detection. 

\textbf{Acknowledgments.} \textcolor{black}{The author is indebted
}to the members of the Laboratory of Theoretical Physics at Magurele-Bucharest
for many fruitful discussions. This work has been supported by the
Scientific Research Agency of the Romanian Government through Grant
PN 19060101/2019

\textbf{Statement:} The author declares no conflict of interest.

\end{document}